\documentclass{PoS}

\usepackage{graphicx}
\usepackage{amssymb,amsmath}

\newcommand{\be}{\begin{equation}}
\newcommand{\ee}{\end{equation}}
\newcommand{\beq}{\begin{equation}}
\newcommand{\eeq}{\end{equation}}
\newcommand{\bea}{\begin{eqnarray}}
\newcommand{\eea}{\end{eqnarray}}

\newcommand{\nn}{\nonumber}
\newcommand{\e}{\hbox{e}\,}
\newcommand{\tr}{\hbox{Tr}}
\newcommand{\Tr}{\textrm{Tr}}

\newcommand{\bra}{\langle}
\newcommand{\ket}{\rangle}

\title{Hopping parameter expansion to all orders using the Complex Langevin equation}

\ShortTitle{Hopping parameter expansion using the Complex Langevin equation }

\author{Gert Aarts\\
        Department of Physics, College of Science, Swansea University,
Swansea SA2 8PP, United Kingdom \\
        E-mail: \email{g.aarts@swansea.ac.uk}}

\author{Erhard Seiler\\
        Max-Planck-Institut f\"ur Physik (Werner-Heisenberg-Institut)
M\"unchen, Germany  \\
        E-mail: \email{ehs@mpp.mpg.de}}

\author{\speaker{D\'enes Sexty} \\
        Bergische Universit\"at Wuppertal\\
        E-mail: \email{sexty@uni-wuppertal.de}}

\author{Ion-Olimpiu Stamatescu\\
        Institut f\"ur Theoretische Physik, Universit\"at Heidelberg, Heidelberg, Germany \\
        E-mail: \email{i.o.stamatescu@thphys.uni-heidelberg.de}}

\abstract{ 
We propose two novel formulations of the hopping parameter 
expansion for finite density QCD using Wilson fermions, while keeping 
the gauge action intact.
We use the complex Langevin equation to circumvent the sign problem in 
the theory. We perform simulations at very high order of the expansion,
such that convergence is directly observable. We compare results to 
the full QCD results, and see agreement at sufficiently high orders. 
These results provide support for the use of complex Langevin dynamics to study QCD at nonzero density, both in the full and the expanded theory, and for the convergence of the latter.
    }

\FullConference{9th International Workshop on Critical Point and Onset of Deconfinement\\
		 17-21 November, 2014\\
		 ZiF (Center of Interdisciplinary Research), University of Bielefeld, Germany}

\begin{document}

\section{Introduction}
 Non-perturbative calculations of the phases
strongly interacting matter
are hampered by the sign problem (for a review of different approaches see 
\cite{signreview}).

In the past years, however, 
 progress in complex Langevin (CL) dynamics 
\cite{Aarts:2009uq,Seiler:2012wz,Sexty:2013ica,Aarts:2014fsa} has 
led to the hope that 
the full region of physical interest can be explored.
In this study we support the complex Langevin simulations 
by exploring an alternative approach to lattice fermions at nonzero
chemical potential: the hopping
parameter expansion, which can be formulated as a systematic approximation
to finite density QCD. The expansion is expected to converge at not too 
small quark masses.

Historically the hopping expansion was used in the form of the 
loop expansion (described in detail in Sec. 3). It was used by 
earlier studies at LO and NLO 
\cite{Bender:1992gn,Blum:1995cb,Bakeyev:2000wm,Aarts:2002} with 
the full Yang-Mills action, also to 
map the phase diagram in \cite{DePietri:2007ak,ben}. In studies combined with 
strong coupling expansion it has been possible to calculate NNLO
contributions as well \cite{Frommpapers,Greensite:2013vza,Langelage}, but it has proven quite difficult to 
further extend the the expansion to higher orders.

Here we present an alternative way to introduce higher-order corrections 
in the hopping parameter expansion \cite{Aarts:2014bwa}. The approach allows 
calculations at 
very high orders (only limited by available computer power), while keeping 
the full Yang-Mills action, and without having to consider fermionic loops 
and their combinatorial factors at each new order.

We define the $\kappa$- and $\kappa_s$-expansions below, with slightly 
different properties. We improve on the convergence properties 
of the loop expansion where the effective expansion parameter is 
$ \kappa N_\tau$ with $N_\tau$ the temporal extent of the lattice.

In section 2, we briefly describe Complex Langevin simulations. In Section 3,
we first review the loop expansion, then describe the new approaches 
we call $\kappa$- and $\kappa_s$-expansion, and discuss their implementation
in the complex Langevin equation. In Section 4 we present 
numerical results gained using this approach. Finally, in Section 5 
we conclude.

\section{Complex Langevin Simulations}

The Complex Langevin approach is based on the complexification
of the Langevin equation \cite{parisi,klauder}. This also leads to 
the complexification of the field manifold. The resulting 
process is susceptible to numerical problems 
(runaway trajectories, solved by using adaptive step sizes \cite{Aarts:2009dg}), 
as well as convergence to a wrong result. Recently it has 
been shown that convergence is guaranteed as long as some conditions 
are satisfied, such as the fast decay of field distributions and 
holomorphy of the action and the observables \cite{Aarts:2009uq}.
Note that there are several types of modifications possible to adapt the 
Langevin process for a given action, which one can 
use to get the process to satisfy convergence criteria \cite{Aarts:2012ft}.
The method has proven useful in other contexts with a complex 
action as well \cite{bosegas,thetaterm,realtime,Mollgaard:2014mga,Hayata:2014kra}.

In lattice QCD the discretised Langevin equation is written as \cite{batrouni}
\bea \label{laneq}
U_{x,\nu}\mapsto \exp\left\{\sum_a i\lambda_a (\epsilon 
K_{x\nu a}+\sqrt{\epsilon}\eta_{x\nu a})\right\}U_{x,\nu}, \;\;
\label{dyn}
\eea
where $K_{x\nu a}= -D_{x\nu a}S $ is the drift force, $\epsilon$ the 
(adaptive) stepsize, and $\eta$ independent Gaussian noises satisfying
$\bra \eta_{x\nu a} \eta_{x'\nu'a'}\ket =2 \delta_{aa'} \delta_{xx'} \delta_{\nu\nu'}$.
 A complex action leads to a complex drift $K$, and link variables  
take values in SL(3,$\mathbb{C}$), losing their unitarity.

 The available configuration space is thus complexified, and loses compactness.
For gauge theories this leads to an additional complication: the volume
of gauge orbits corresponding to a configuration is infinite. To restrict 
the movement of the system along the infinite gauge orbits one 
has to modify the process, while respecting the gauge invariance of the 
action and observables. This can be very conveniently achieved 
with the gauge cooling  \cite{Seiler:2012wz} (see also the review \cite{Aarts:2013uxa}), which uses non-compact gauge transformations to force the process to stay near the unitary manifold. As a consequence the 
decay of the distributions is fast, as required for the convergence proof.
Together with the adaptive stepsize this practically eliminates runaways.

Another requirement for the proof of convergence is the 
holomorphy of the action, which is unfortunately not satisfied 
for QCD. This manifests in zeros of the measure, i.e.\ $\det
M=0$, leading to a meromorphic drift. 
Poles in the drift then might lead to 
wrong convergence of the process, as shown in nontrivial, soluble
models \cite{Mollgaard:2013qra}, while in many cases the process
gives correct results in spite of a non-holomorphic action, especially in the 
cases where the non-holomorphy is due to a Haar measure or Jacobian 
\cite{Aarts:2012ft}.

\section{Hopping parameter expansions}

\subsection{Loop expansion}

Recall the path integral formulation of QCD,
where we use the plaquette action $S_{\rm YM}$ for the gauge fields 
\be
Z = \int DU\, \e^{-S},\quad S= S_{\rm YM}- \log \det M 
\label{e.zqcd},
\ee
with the Wilson fermion matrix $M$, the hopping term $Q$ of which we split into 
spatial hopping terms $S$ and temporal hopping terms $R$
\be
M = 1-\kappa Q = 1-\kappa_s S - R,
\label{eqMQCD}
\ee
with
\bea
 S_{xy} &=& 2\sum_{i=1}^3  \left( \Gamma_{-i} U_{x,i} \delta_{y,x+a_i}  
+ \Gamma_{+i} U_{y,i}^{-1} \delta_{y,x-a_i}  \right), \\
 R_{xy} &=&  2\kappa\left( e^\mu \Gamma_{-4} U_{x,4} \delta_{y,x+a_4}  
+ e^{-\mu} \Gamma_{+4} U_{y,4}^{-1} \delta_{y,x-a_4}\right), \nn
\eea
using the matrices $\Gamma_{\pm \nu} = (1 \pm \gamma_\nu)/2$. Note that 
these matrices are projectors satisfying $\Gamma_{\pm \nu}^2 = \Gamma_{\pm \nu}$ and $\Gamma_{+\nu}\Gamma_{ -\nu} =0$.
We then expand the fermionic part of the measure
\bea
\det M = 
\exp ( \Tr \log (1 - \kappa Q) ) = \exp \left( - \tr \sum_{n=1}^\infty 
{ \kappa^n \over n} Q ^n  \right), 
\eea 
which we can rewrite noticing that we can perform the sum for each 
loop built from hopping terms on the lattice separately, then resum the 
determinant for each loop (ignoring possible convergence problems)
\bea
\det M =  \exp  \left( - \tr \sum_{C,s=1}^\infty 
{ \kappa^{l_C s} \over s} L_C ^s  \right) = \prod_C \det ( 1 - \kappa^{l_C} L_c ),
\eea
where $C$ goes over all the possible loops on the lattice and $l_C$ is 
the length of the loop. Since $ \Gamma_{+\nu} \Gamma_{-\nu}=0 $, no 
loop that turns back on itself needs to be considered.

\begin{figure}[t]
\begin{center}
\includegraphics[width=0.48\columnwidth]{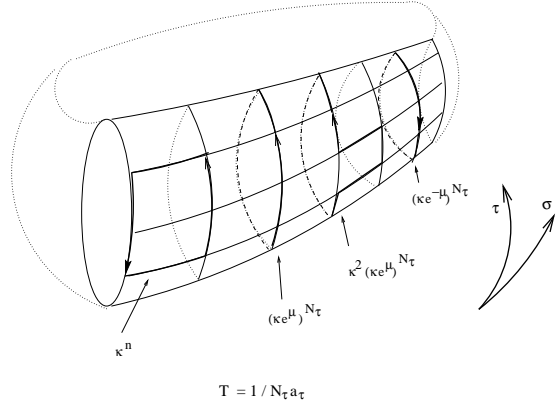}  
 \caption{ An illustration of the different loops contributing 
to the loop expansion of the fermion determinant. }
\label{torus}
\end{center}
\end{figure}

In the static limit, where $\kappa \rightarrow 0, \ \mu \rightarrow \infty,
\ \zeta= 2 \kappa \exp{ \mu} = \textrm{const.}$, 
all contributions vanish except for the Polyakov loops. This is called the 
leading order of the loop expansion. Next to leading order (NLO) loops 
are gained by a decorating the Polyakov loop with two spatial hoppings,
as illustrated in Fig.~\ref{torus}.
NNLO contributions involve Polyakov loops with more decorations as well as 
the plaquette contribution.

\subsection{$\kappa$ expansion and $\kappa_s$ expansion}

To define an expansion which can be conveniently included 
in the Langevin equation, we go back to the first step in the loop expansion
\bea \label{kappaexp}
\det M = 
\exp ( \Tr \log (1-\kappa Q )) = \exp \left( - \tr \sum_{n=1}^\infty 
{ \kappa^n \over n} Q ^n  \right) .
\eea 
In the sum only even powers of the hopping matrix contribute, as a trace 
 is present. The expression is straightforwardly 
generalized to $N_f>1$ flavors. We call this expansion the $\kappa$-expansion.
The fermionic observables can than be expressed in 
terms of the expansion as
 \bea
  \bra \bar \psi \psi \ket &=& \frac{2\kappa N_f}{\Omega} \label{meas-cc}
  \sum_{n=0}^\infty \kappa^n  \left\bra \tr\, Q^n \right\ket, \\
 \bra n\ket &=& -\frac{N_f}{\Omega} \sum_{n=1}^\infty \kappa^n \left\bra
 \tr \left({ \partial Q \over \partial \mu} Q^{n-1}\right) \right\ket, \label{meas-dens}
\eea 
for the chiral condensate $\bra \bar \psi \psi \ket$ and baryonic density $n$, using $\Omega=N_s^3N_\tau$ the lattice volume.

Alternatively, we can rewrite the fermion matrix using the identity
\be
M = (1-R) \left( 1 - \frac{1}{1-R} \kappa_s  S \right).
\ee
before expanding to gain the following expansion:
\be
\det M = \det (1-R) \exp \sum\limits_{n=1}^{\infty}  
- \frac{\kappa_s^n}{n} \tr\left(  \frac{1}{1-R}  S \right)^n.
 \label{kappasdet}
\ee
Since the matrix $(1-R)^{-1}$ is diagonal in spatial indices 
(and dense in temporal indices) and $S$ describes spatial hoppings,
again only even powers contribute in the sum. The first factor 
of the expansion $\det(1-R)$ describes the LO contribution in the 
static limit, also known as the HDQCD approximation. This simplification
happens only for Wilson fermions, as backtracking is forbidden, hence the 
only possible loop built from temporal hoppings is the Polyakov loop.
In this case the determinant and the inverse of the matrix can be 
calculated analytically as follows. The inverse of the matrix 
can be written as 
\bea
(1-R)^{-1} =  (1-R^+)^{-1} + (1-R^-)^{-1} -1
\eea
with $R^+ + R^- = R,\ R^+ R^- = 0 $ and $R^+$ ($R^-$) containing hoppings in 
positive (negative) temporal directions. We can then find the inverse of the 
two terms by the simple expansion (omitting spatial coordinates)
\bea
(1-R^+)^{-1}_{xy}= \sum_{n=0}^{\infty}\left(2\kappa e^\mu \Gamma_{-4} 
U_{x,4} \delta_{y,x+a_4}\right)^n,
\label{eq:10}
\eea
 Separating the parallel transporter
 between $x$ and $y$, we can easily resum the remaining factor to give
\bea \label{1mRinverz}
(1-R^+)^{-1}_{xy} &=&  1  -\Gamma_{-4}  { (2 \kappa e^\mu)^{N_\tau} P(x)
 \over 1 +  (2 \kappa e^\mu)^{N_\tau} P(x) } 
\quad \textrm{if } x=y \\ \nonumber 
 &=&  \Gamma_{-4} (2 \kappa e^\mu)^{y-x} { 1 
 \over 1 +  (2 \kappa e^\mu)^{N_\tau} P(x) } W(x,y)
\quad \textrm{if } y>x \\ \nonumber
&=& -\Gamma_{-4} (2 \kappa e^\mu)^{N_\tau+y-x} { 1 
 \over 1 +  (2 \kappa e^\mu)^{N_\tau} P(x) } W(x,y)
\quad \textrm{if } y<x \\ \nonumber 
\eea 
where $W(x,y)$ is the parallel transporter between $x$ and $y$ 
built from positive hoppings, $P(x)$ is the untraced Polyakov loop 
starting from site $x$ (that is $P(x)=W(x,x)$). 
The inverse of $(1-R^{-})$ is calculated similarly.
 Similar formulas were derived in 
Refs.\ \cite{Langelage}  
in an effort to develop an effective theory for Polyakov-loops,
also utilizing the strong coupling expansion for the Yang-Mills action.

The observables in the $\kappa_s$ expansion can be recovered using the 
defining equations such as $ \bra n \ket = \partial \ln Z / \partial \mu $
in straightforward calculations to yield formulas similar
to (\ref{meas-cc}).

These two expansions can be very conveniently implemented 
in Langevin simulations, as detailed in the next subsection, but they have 
different strengths and weaknesses. We consider a truncated version
of the expansion to order $N^qLO$, in which terms up to $\kappa^{2q}$
are kept. We keep also the terms proportional to $ \exp(-\mu)$, although their
contribution is suppressed at large $\mu$, but they lead to the determinant
satisfying the symmetry:
\bea
\det M(\mu) = ( \det ( M ( -\mu^*) )^ *  
\eea

The $\kappa$ is expansion is very 
cheap to calculate, but its convergence properties at nonzero 
chemical potential are not optimal, as $Q$ has terms proportional 
to $ \kappa \exp(\mu)$. The action truncated to some order 
is holomorphic, so proofs
of convergence (requiring also fast decaying distributions) apply 
\cite{Aarts:2009uq}. In the $\kappa$ expansion
one needs to go to order $\kappa^{4}$ in order to see any $\kappa$ dependence,
as the first closed loop (the plaquette) can be formed at this order. 
Similarly, one needs to go to order $\kappa^{N_\tau}$ in order to see $\mu$ 
dependence (using the Polyakov loop), since for shorter loops 
the dependence cancels. As we will 
demonstrate below calculating high orders is easy in this setup, so this 
drawback is not a serious one.

In our second scheme, the $\kappa_s$-expansion the main part of the 
$\mu$ dependence is dealt with analytically, therefore one expects better
convergence properties at high $\mu$, and this is indeed satisfied,
see in Section. \ref{resultssec}. The price to pay is the slightly higher 
numerical cost and the non-holomorphic action.

\subsection{$\kappa$- and $\kappa_s$ expansion in Langevin simulations}

It is useful to consider the expanded determinant (\ref{kappaexp})
as part of the action. Note that the resulting effective action is holomorphic.
It has than a contribution to the drift term of the Langevin equation 
(\ref{laneq})
\bea
K_{x\nu a}&=&  -\sum_{n=1}^\infty \kappa^n\tr \left( Q^{n-1} 
D_{x\nu a} Q \right).
\eea
Note that this contribution is non-real, therefore we have to complexify 
the theory and use complex Langevin dynamics. (At $\mu=0$ real Langevin 
simulations are possible by taking the real part of the fermionic drift terms.)

This contribution to the drift term can be estimated using a random vector  $ \eta_i$  
(where  $i$ represents space-time, colour and Dirac indices),
with the properties $\bra \eta_i \ket=0$, $\bra \eta^*_i \eta_j 
\ket= \delta_{ij}$ as
\bea \label{noisyforce}
K_{x\nu a}= \bra\eta^* (D_{x\nu a} Q)  s\ket, \quad\quad
  s = -\sum_n \kappa^n Q^{n-1} \eta. 
\eea
 The calculation of this term at $N^qLO$ thus requires $2q$ multiplications 
with the sparse matrix $Q$ in every timestep, when the random vector is 
refreshed.
 In the case of the $\kappa_s$-expansion, the drift term has contributions 
from several places. The $ \ln \det (1-R)$ term has contributions to 
the drift identical to the HDQCD, as calculated in \cite{Aarts:2008rr}.
 The contribution of the expansion is  
\bea
K_{xia} &=&  - \sum\limits_{n=1}^\infty  \kappa_s^n 
\tr \left({1\over 1-R} (D_{xia} S)  
 \left[ {1 \over 1-R}  S \right]^{n-1} \right), \nn \\
K_{x4a} &=&  - \sum\limits_{n=1}^\infty \kappa_s^n 
\tr \left( {1 \over 1-R } (D_{x4a} R)  
 \left[  {1 \over 1-R}  S \right]^{n} \right), \;\;\;\; 
\eea
for spatial and temporal links, correspondingly. These contributions are
 estimated using noise vectors similarly to (\ref{noisyforce}). The numerical
effort of the $\kappa_s$ expansion involves also multiplications 
with $ (1-R)^{-1}$, calculated according to (\ref{1mRinverz}), as well as 
multiplications with the sparse $S$ matrix.

\begin{figure}[h]
\begin{center}
\includegraphics[width=0.70\columnwidth]{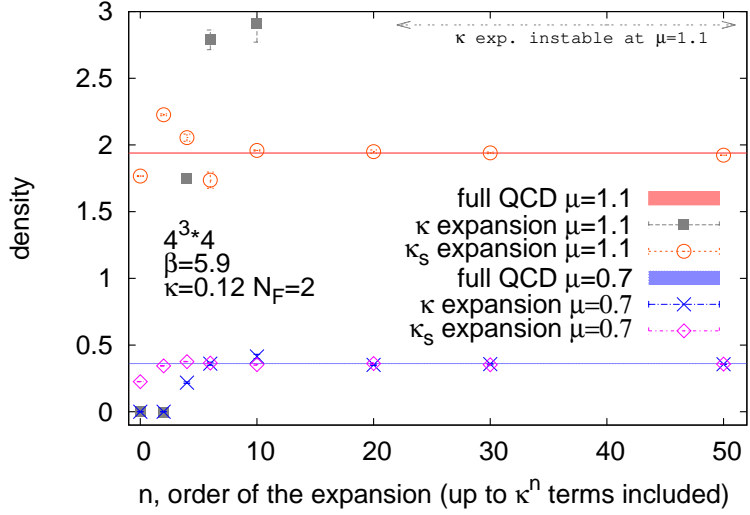}  
\caption{Dependence of the quark density (in lattice units) on the order of the truncation of the $\kappa$- and $\kappa_s$-expansions, for $\mu=0.7$ and $1.1$, on a $4^4$ lattice with $\beta=5.9$, $\kappa=0.12$, and $N_f=2$. The region where the $\kappa$-expansion breaks down for $\mu=1.1$ is indicated.
The lines show the result for full QCD.
Saturation density is $n_{\rm sat}=2N_cN_f=12$.
}
\label{f_nlopp_dens}
\end{center}
\end{figure}

\section{ Numerical results}
\label{resultssec}

We have used simulations on small $4^4$ lattices to examine the convergence 
properties of the expansions. We have used two flavors of Wilson fermions 
with $\beta=5.9$ and 
$\kappa=\kappa_s=0.12$, for several $\mu$ values. 
We compare results with full QCD, obtained by complex Langevin simulations,
extending the previous results available for 
staggered fermions  \cite{Sexty:2013ica} to Wilson fermions.
Since the full QCD result is also obtained with Wilson fermions
there is no need to renormalize the results, already the bare quantities 
of the expansions in lattice units should converge to their full QCD values.
We plot results
in lattice units. The lattice spacing is measured using the 
gradient flow, as proposed in Ref.\ \cite{Borsanyi:2012zs}. 
The lattice spacing depends weakly on the mobility of the fermions for 
the heavy quark masses that we are using. For HDQCD we find that $\beta=5.9$ and $\kappa=0.12$ corresponds to  $a \simeq 0.12$ fm, while for full QCD we find $a \simeq 0.114$ fm.

In Fig.~\ref{f_nlopp_dens} we show the quark number density as a function 
of the order of the $\kappa$- and $\kappa_s$-expansions
for two different $\mu$ values, comparing to the full QCD result.
We see that at the smaller chemical potential $\mu=0.7$ both expansions 
behave similarly, with convergence around the $\kappa^{10} $ order.
At the higher chemical potential value, the $\kappa$ expansion breaks down, as 
expected. 

We observe good convergence of the series to the full QCD values.
This is a non-trivial agreement which supports both the expansion and 
the full QCD. 
This means in particular in the case of the $\kappa$ expansion, that 
we must obtain the correct value also in full QCD. This apparently means 
that the non-holomorphicity of the action for full QCD is not a problem
(at least for the case where the $\kappa$ expansion converges).

\begin{figure}[h]
\begin{center}
\includegraphics[width=0.48\columnwidth]{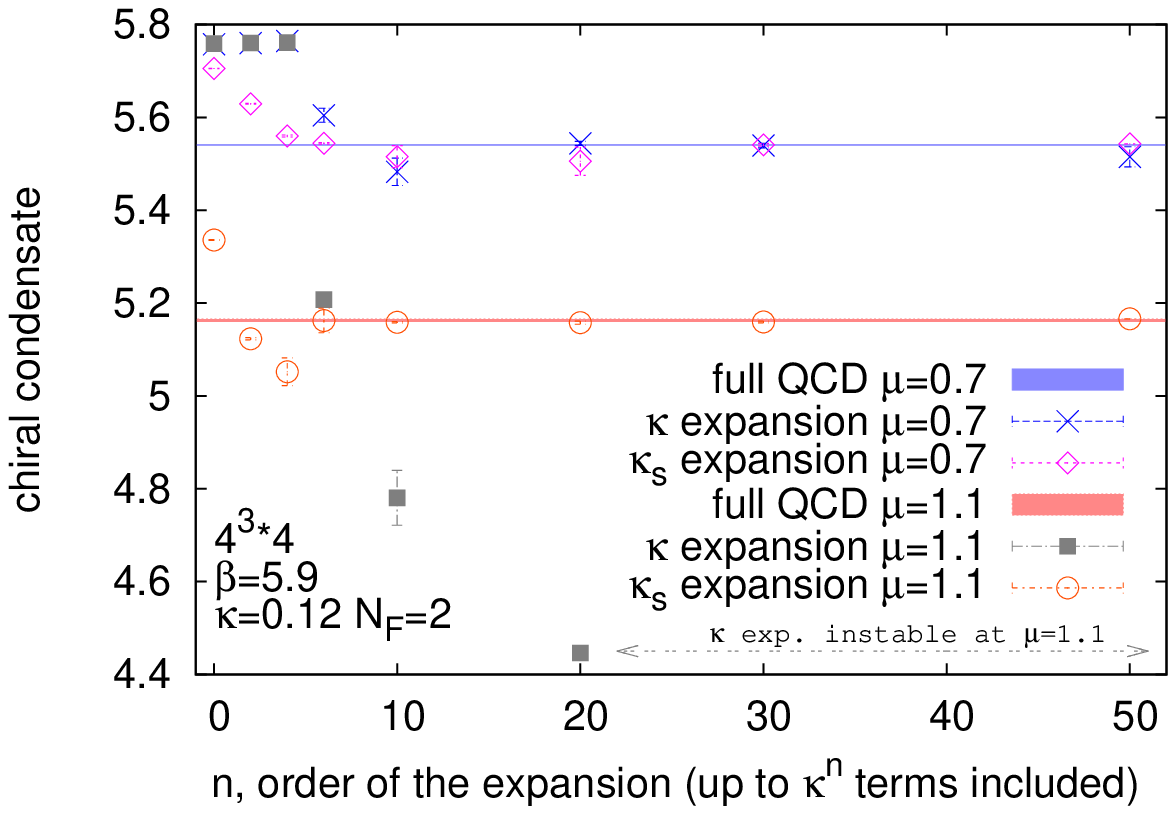}  
\includegraphics[width=0.48\columnwidth]{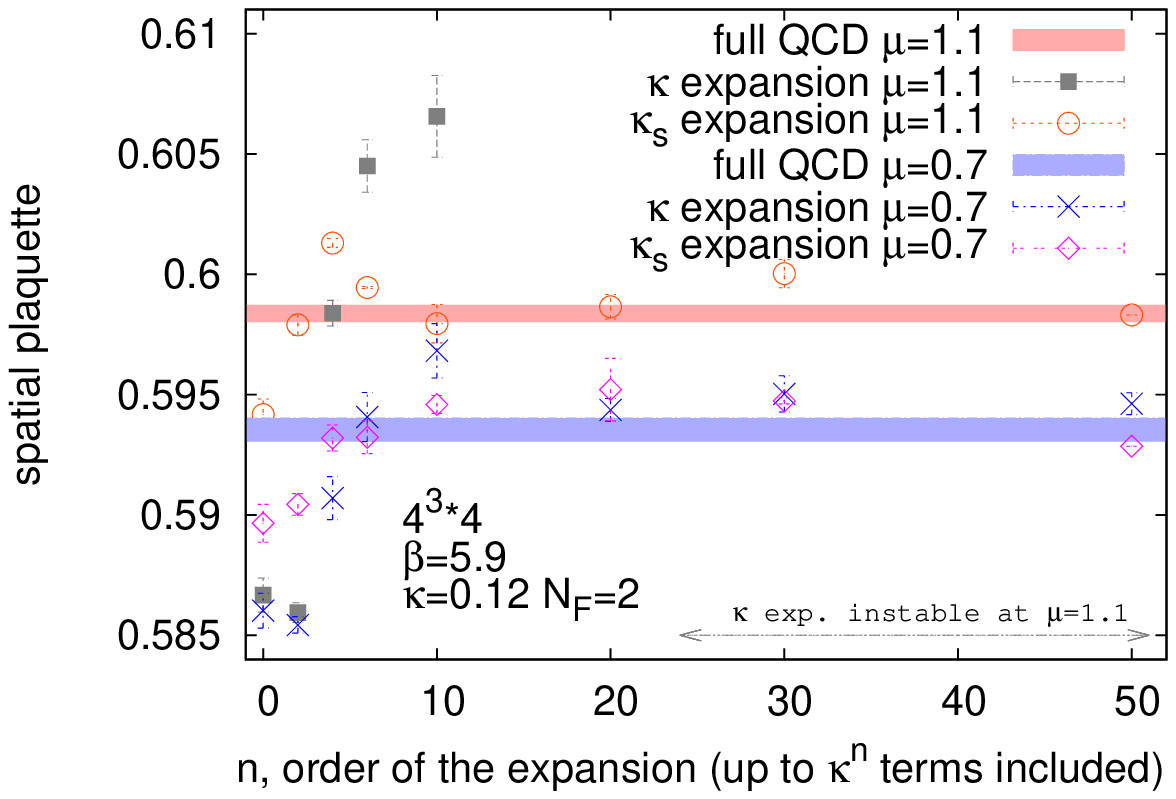}  
\caption{Dependence of the chiral condensate and the spatial plaquette average (in lattice units) on the order of the truncation of the $\kappa$- and $\kappa_s$-expansions, for $\mu=0.7$ and $1.1$, on a $4^4$ lattice with $\beta=5.9$, $\kappa=0.12$, and $N_f=2$. The region where the $\kappa$-expansion breaks down for $\mu=1.1$ is indicated.
The lines show the result for full QCD.
}
\label{f_nlopp_ccplaq}
\end{center}
\end{figure}

In Fig.~\ref{f_nlopp_ccplaq} we see similar behavior for the chiral condensate and the spatial plaquette average.

\begin{figure}[t]
\begin{center}
\includegraphics[width=0.48\columnwidth]{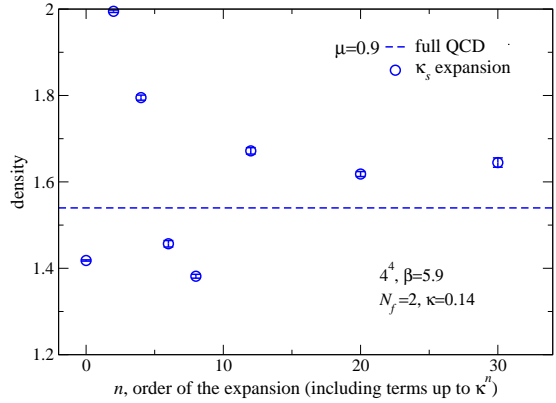}  
 \caption{
The density at $\kappa=0.14$ on a $4^4$ lattice for the $\kappa_s$-expansion and full QCD. }
\label{fig:nemconv}
\end{center}
\end{figure}

\begin{figure}[t]
\begin{center}
\includegraphics[width=0.58\columnwidth]{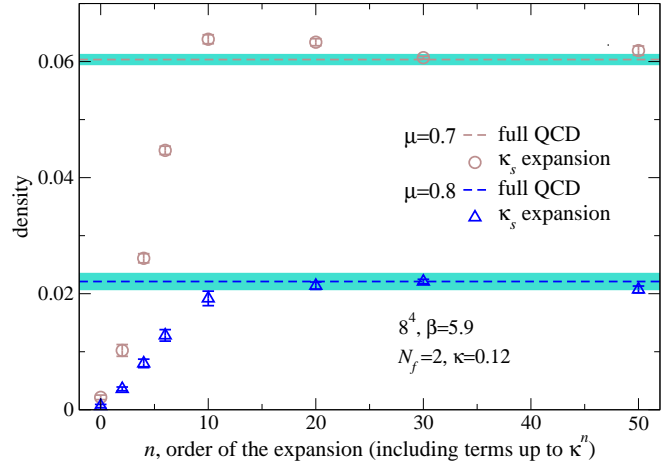}  
 \caption{ 
The density at $\kappa=0.12$ on a $8^4$ lattice for the $\kappa_s$-expansion and full QCD for chemical potential $\mu=0.7$ and $\mu=0.8$. }
\label{fig:8conv}
\end{center}
\end{figure}

While the convergence appears quite quick at $\kappa=0.12$, 
at larger $\kappa$ this might not be the case, see 
Fig.~\ref{fig:nemconv} where we show the 
performance of the $\kappa_s$ expansion 
at $\kappa=0.14$. These results suggest that the convergence radius of the 
$\kappa_s$ expansion is below $\kappa=0.14$ at $\mu=0.9$. 
The convergence radius seems to be independent of the lattice size, however.
In Fig.~\ref{fig:8conv} we show the convergence of the density on a $8^4$ lattice.
This lattice system has a temperature below the deconfinement transition. 
We see that density grows about a factor of 3 as one changes the chemical
potential from $\mu=0.7$ to $\mu=0.8$, which is a sign of the 
rapid onset transition.  One observes that the $\kappa_s$ 
expansion still performs well in this region.

\section{Conclusions}

 In this study we have presented the novel implementations of the 
hopping expansion for the complex Langevin equation which are called
$\kappa$- and $\kappa_s$ expansion. They allow calculations 
at very high, previously impossible orders. This allows the direct 
observation of the convergence of the series to the full QCD result.
We use the complex Langevin equation to circumvent the sign problem
of these theories at finite chemical potential. We use no further 
approximation other than the hopping expansion, the gauge action 
is kept intact in particular, thus our method can also be used as a test ground 
for other effective models. 

Our expansions have different merits: the $\kappa$ expansion is cheap and has 
a holomorphic action, but its convergence properties are bad at 
large chemical potentials. The $\kappa_s$ expansion is slightly more
expensive numerically, but has improved convergence properties also 
at high chemical potentials.

We performed simulations of the expansions and observed good convergence to 
full QCD at not too high $\kappa$ parameters. This convergence supports 
both the expanded and the full theory, as the agreement shows that 
the non-holomorphy of the full theory has apparently no impact on the 
results, at least in the parameter range where the convergence is observed. 

The first results indicate that at least the onset transition might be within 
the reach of this method in the cold and dense region of the QCD phase diagram,
but further studies are required at low temperatures on large lattices.

\end{document}